\documentclass[aps,prb,showpacs,twocolumn]{revtex4}
\usepackage{amssymb}
\usepackage{amsmath}
\usepackage{graphicx}

\begin{document}

\title{Charge density wave sliding driven by an interplay of conventional and Hall voltages in NbSe$_3$ microbridges}
\date{\today}

\author{A.V.~Frolov$^1$, A.P.~Orlov$^1,2$, A.A.~Sinchenko$^{1,3}$, and P.~Monceau$^{4,5}$}

\address{$^{1}$Kotelnikov Institute of Radioengineering and Electronics of RAS, 125009 Moscow, Russia}

\address{$^{1}$Institute of Nanotechnologies of Microelectronics of RAS, 115487, Moscow, Russia}

\address{$^{3}$M.V. Lomonosov Moscow State University, 119991, Moscow, Russia}

\address{$^{4}$Univ. Grenoble Alpes, Inst. Neel, F-38042 Grenoble, France, CNRS, Inst. Neel, F-38042 Grenoble, France}

\address{$^{5}$Univ. Grenoble Alpes, INSA Toulouse, Univ. Toulouse Paul Sabatier, CNRS LNCMI, 38000 Grenoble, France}

\begin{abstract}
	
Collective charge-density wave (CDW) transport was measured under a high magnetic field in NbSe$_3$ microbridges which have been cut transversely and at an angle to the chains' direction. We give evidences that the CDW sliding is driven by the Hall voltage generated by the inter-chain current of normal carriers. We have discovered a re-entrance effect of the Hall-driven sliding above a crossover temperature at which the Hall constant has been known to change sign. For the narrow channel, cut at 45$^\circ$ relative to the chain axis, we observed an evolution from the Hall-driven sliding at low temperatures, to the conventional sliding at higher temperatures, which corroborates with falling of the Hall constant. In this course, the nonlinear contribution to the conductivity coming from the collective sliding changes sign. The quantization of Shapiro-steps, generated presumably by a coherent sequence of phase slips, indicates that their governing changes from the applied voltage to the current.

\end{abstract}


\maketitle

\section{Introduction}\label{Intr}

The collective motion (sliding) of charge density waves (CDW) is one of the most prominent properties of low-dimensional
compounds exhibiting this type of electronic ordering. First predicted by Fr\"{o}hlich \cite{Frohlich54} as dissipationless
electron transport, this sliding is possible only when an electric field exceeds a certain
characteristic threshold electric field, $E_t$ \cite{Fleming79}. The sliding of the charge density wave,
which is manifested as a sharp {\it increase of conductivity} in fields $E>E_t$, was previously observed and well
studied in many quasi-one-dimensional compounds \cite{Gruner,Monceau12}. Fr\"{o}hlich suggested that there can be states with current flow if the CDW energy gap is displaced with the electrons and remains attached to the Fermi surface. This model was revisited in Refs.(\onlinecite{Bardeen74,Bardeen85,Bardeen89}). If the electron dispersion is displaced by $q$, that leads to a Fr\"{o}hlich collective current $j=nev_s$ with $\hbar q=m^*v_s$, where $m^*$ is the Fr\"{o}hlich CDW mass. 

\begin{figure}[t]
	\includegraphics[width=8.5cm]{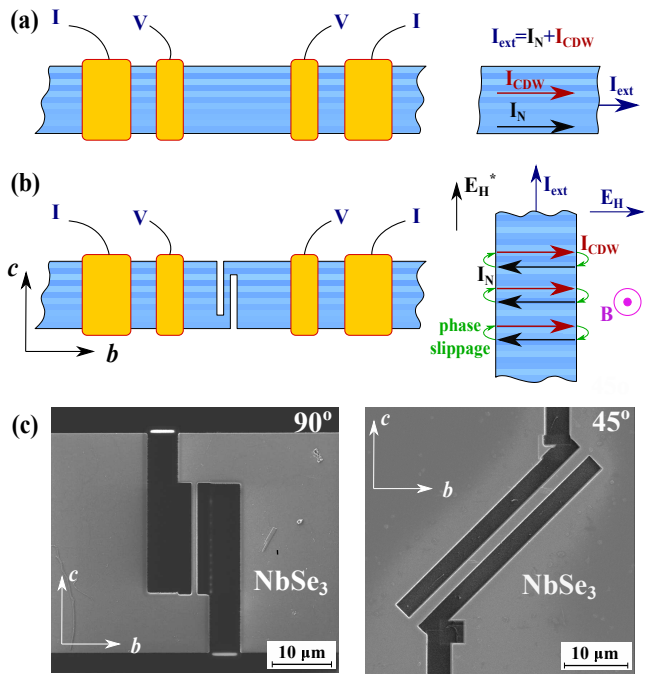}
	\caption{Scheme of devices with current and voltage distribution: a) in the conventional configuration; b) in the Hall configuration; c)   images of NbSe$_3$ structures: left - 90$^\circ$ microbridge cut along $c$-axis transverse to the CDW chains ($b$-axis); right - microbridge oriented under 45$^\circ$ degree relative to the $b$-axis. In the black rectangles, NbSe$_3$ was lifted off by focusing ion beam (FIB).} \label{F0}
\end{figure}

In conventional experiments on CDWs, the driving force for the depinning and the subsequent sliding is provided by the
electric field, $E$, oriented along the direction of the CDW chains (Fig. \ref{F0}a). Commonly, the penetration of $E$ is originated by the applied voltage, $V$, and mediated by the current $j_n$ of remnant normal carriers. The collective CDW current $j_c$ sets in above a threshold $j_n>j_t$ and both currents $j_n$ and the $j_c$ flow in the same direction. In the non-linear state a periodically time dependent voltage is generated \cite{Thorne88}. Interference effect between these voltage oscillations and an external radio-frequency electric field is manifested in Shapiro steps which appear as a sequence of peaks in the $dV/dI(I)$ characteristics which are equidistant in the current \cite{Zettl84}.

Recently, non-linear transport properties were measured with the current applied perpendicular to the chain axis of NbSe$_3$. NbSe$_3$ is a prototype material which undergoes two successive incommensurate transitions at $T_{P1}=144$ K and $T_{p2}=59$ K. NbSe$_3$ crystallises in a ribbon-like shape with the chain axis along $b$, the ribbon being parallel to the ($b,c$) plane. The structure is monoclinic with space group $P_{2_1}/m$.

A novel nonlinearity of the threshold type in the electronic transport of NbSe$_3$ was observed in narrow bridge-type structures cut along the $c$-axis across the CDW chains under a magnetic field directed in the another transverse direction $B\parallel a^*$. The scheme of the device as well as the current and the electric field configuration are shown in Fig. \ref{F0}. Above some threshold current $j_t$ the {\it conductivity sharply drops down} by several times. The non-linear regime is accompanied by the generation of a coherent electromagnetic radiation observed up to the GHz range; that was identified by the observation of Shapiro step type response in the current-voltage characteristics (IVc) to an external radiation. This effect was interpreted as the CDW being driven by the Hall electric field, $E_H$, oriented along the chains $b$-axis (Fig. \ref{F0}b) originated from the mutually orthogonal normal current running along the $c$-axis and the magnetic field along $a^*$-axis. Since the circuit is open along the $b$-axis, the depinning of the collective CDW current at $E_H>E_t$ must be compensated by a counter-current of normal carriers \cite{QM17}. The zero-sum loop of these two currents is closed by periodic phase slip processes \cite{Monceau12} giving rise to spontaneous coherent oscillations. The appearance of the normal counter-current gives rise to a secondary Hall voltage, $V_{H}^*$, now in the channel direction ($c$-axis), the persistent part of which is registered as a sharp threshold in the IV-curve. As a result, the additional drop of voltage, $V_H^*$, which is directly proportional the CDW current is measured in the experiment. In this configuration the CDW sliding is seen (as a mirror effect) by the occurrence of a normal counter-current to ascertain the null current in the channel. 

In this scenario, the CDW is put in motion when the normal current along the CDW chains is absent, both currents appearing simultaneously and, curiously, running in opposite directions, which is completely in contrast to the usual geometry. This effect was best observed at lowest temperatures, weakened with growing $T$ and disappeared completely at $T\sim35$ K. Such a behavior is opposite to CDW sliding in the conventional geometry when the effect of sliding is most pronounced at high $T\sim30-50$ K. At low $T<10$ K this effect is still possible to be observed only using pulse  measurements of current-voltage characteristics (IVc) \cite{Latyshev11}. For explaining these complex results a model was proposed in Ref. \onlinecite{QM17} in which , at low $T$, the remnant normal carriers are in extreme quantum regime and the imposed transverse electric current was put into the regime of a self-tuned nearly integer quantum Hall (IQH) state. This model allowed explaining the observed non-linearity of IVc and high-frequency generation. However, Hall effect exists in NbSe$_3$ well above 35 K \cite{Hall09,Hall08,Coleman90} and the natural question arises: why does the non-linearity of IV-curves disappears about $T\sim35$ K and what does happen at this particular temperature? In this article we shall find a re-entrance effect that the Hall driven sliding emerges back at higher temperatures and can be traced up to $T>50$ K. The particular region of 35K happens to be the one where the Hall voltage is compensated while the Hall constant changing sign.  

In experiments with slit geometries the CDW chains are spatially limited by the width of the microbridge having no direct contact with normal electrodes. This width is very small, typically 1-5 $\mu$m for the structures studied in Ref. \onlinecite{QM17}. The questions arise: is it possible nevertheless to add the direct injection of the normal current, like in conventional sliding geometry, while keeping simultaneously the driving mechanism by the Hall voltage, and what will be the interference of these two mechanisms? 

We shall show indeed that the CDW sliding in samples which are spatially limited along the chain direction is possible not only under Hall electric field but also by the application of an external voltage. By cutting microbridges at 45$^\circ$ relative to the $b$-axis, we succeeded in tracing the transformation of conventional CDW sliding to the Hall driven sliding by varying temperature in a large extent.

\begin{figure*}[t]
	\includegraphics[width=0.3\textwidth,height=5.5cm]{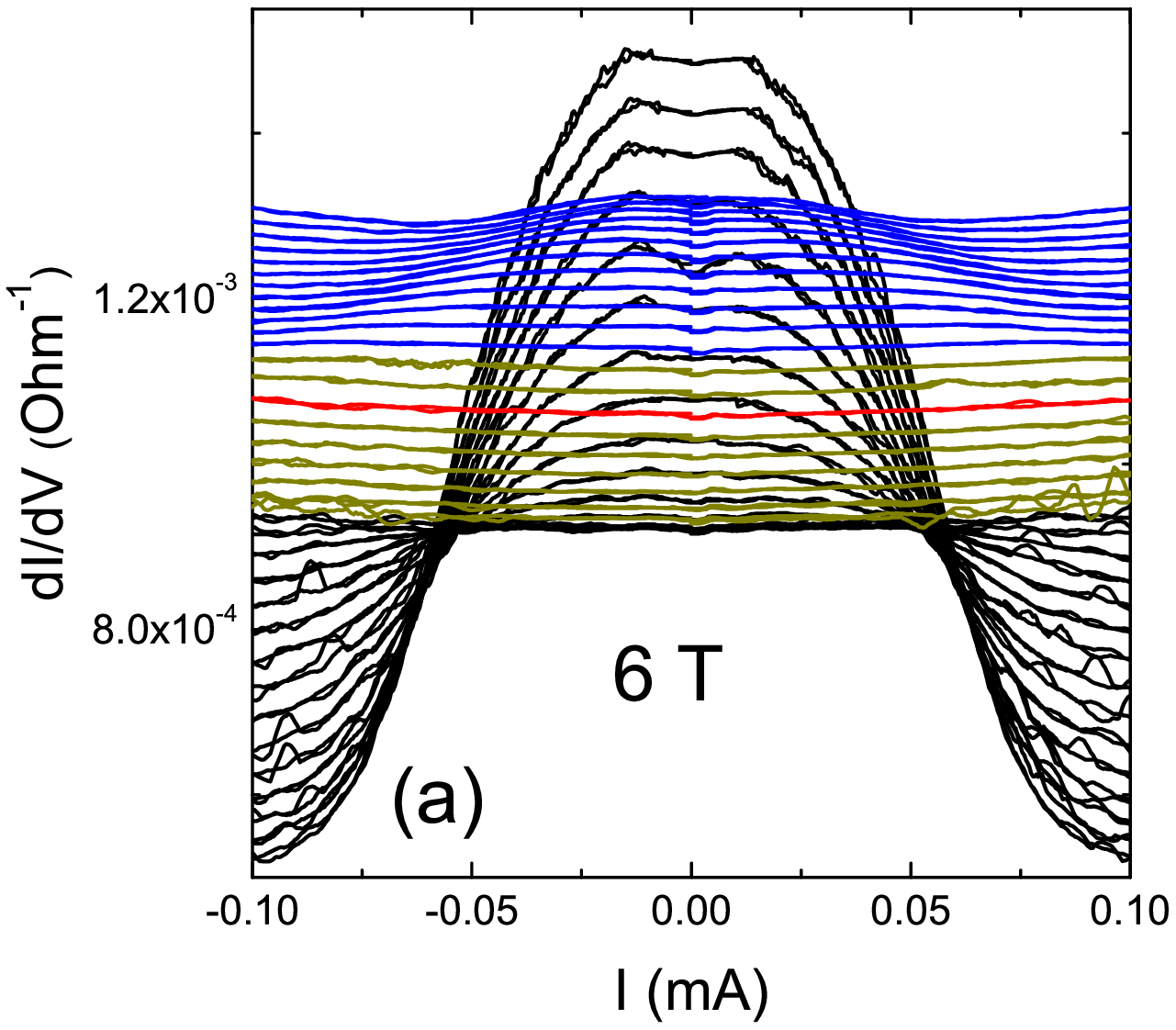} \ \
	\includegraphics[width=0.3\textwidth,height=5.5cm]{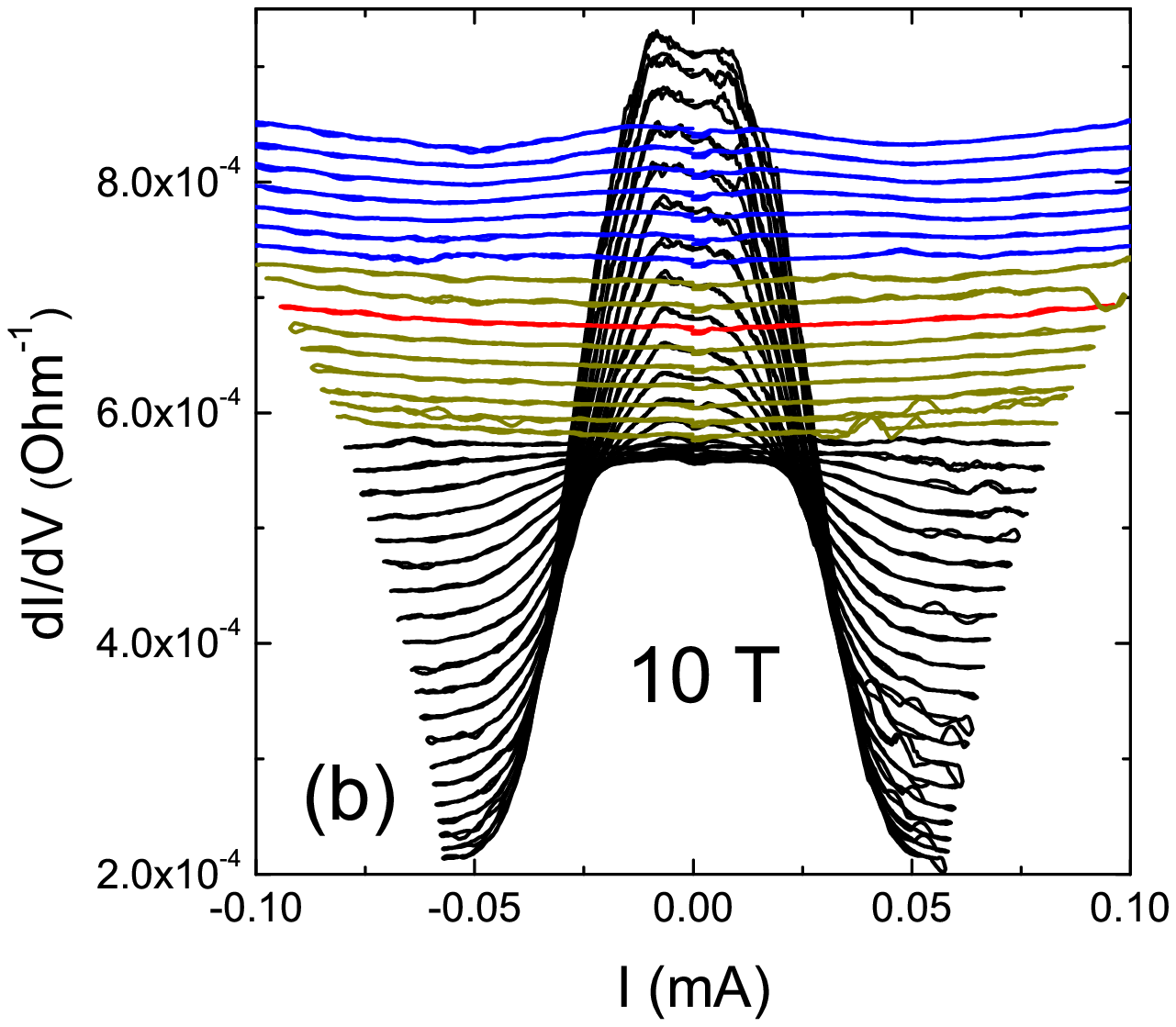} \ \
	\includegraphics[width=0.3\textwidth,height=5.5cm]{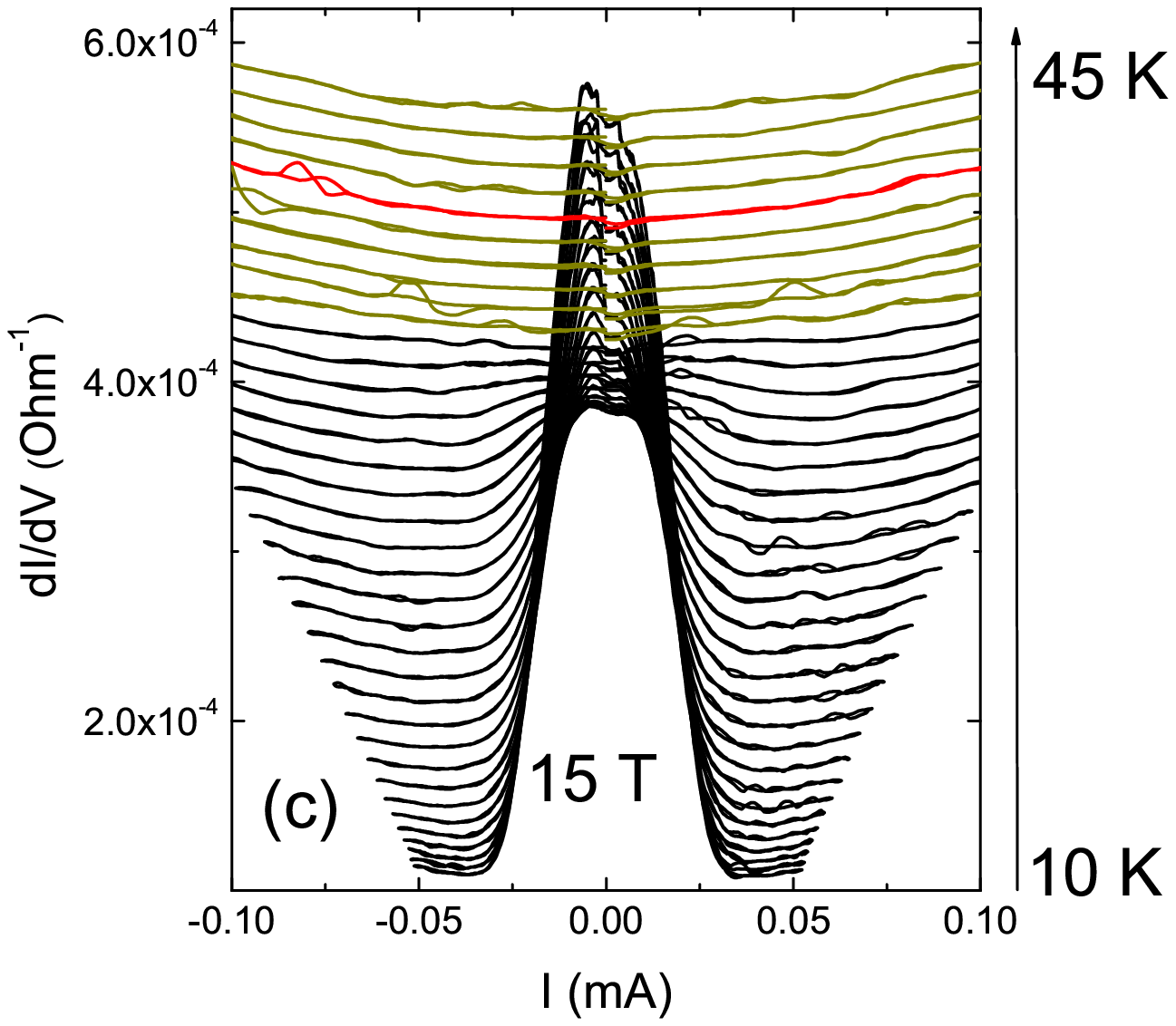}
	\caption{Temperature evolution of differential IVc, $dI/dV(I)$, of a NbSe$_3$ 90$^\circ$ channel in the temperature range 10-45 K in magnetic field $B$: 6 T (a); 10 T (b) and 15 T (c). Distance between neighbour curves: $\Delta T=1$ K. Red curves correspond to the zero-crossing Hall constant. Dark yellow curves indicate the temperature range in which the Hall electric field is less than the threshold electric field for the CDW sliding. Blue curves are nonlinear IVc at high temperatures when a finite Hall voltage is recovered.} \label{F1}
\end{figure*} 

\section{Experimental}

For the experiment we selected thin crystals of NbSe$_3$ of high quality with a thickness less than 1-3 $\mu$m, a width of several tens $\mu$m and a length of several hundred $\mu$m. The crystals were cleaned in oxygen plasma and have been glued to a saphire or crystalline quartz substrate by collodion diluted in amylacetat. For making the structure we used focus ion beam (FIB) machine SMI-3050, Seiko Inc., Japan, with the resolution of Ga-ion beam of 6 nm in the regime of low current of few tens of pA. The electrical contacts to the crystal have been made by pressing indium stabs to golden contact pads prepared by laser evaporation before processing by FIB. The measurements of IV characteristics and their derivatives were carried out in conventional 4-probe configuration using computer controlled current source and nano-voltmeter. Experiments in magnetic field up to $B=20$ T have been carried out in a superconducting magnet in the National Laboratory of High Magnetic Fields in Grenoble.

Two types of structures were prepared and studied (Fig. \ref{F0}): channels with a length 30 $\mu$m and a width 4 $\mu$m oriented across the conducting chains, along the c-axis, which are the same as those studied in Ref. \onlinecite{QM17}, and oblique channels with the same length and width but cut at 45$^\circ$ relative to the $c$-axis (90$^\circ$-cut and 45$^\circ$-cut microbridges or channels in the following). With the current applied to oblique structures, the normal current has a component along both directions, $b$-axis and $c$-axis. In this case, one can expect the interference of two types of sliding: 1) the conventional CDW sliding with collinear normal and collective currents if the component of $E$ along chains ($b$-axis) is larger that $E_t$ and 2) Hall-driven sliding with antiparallel normal and collective currents when the component of the Hall voltage along the $b$-axis overcomes the threshold value. That gives us the possibility of observing and studying simultaneously both types of sliding due to either an applied external voltage or to the Hall voltage in the same structure in which the chain direction is limited by the channel edges.

\begin{figure}[t]
	\includegraphics[width=8.5cm]{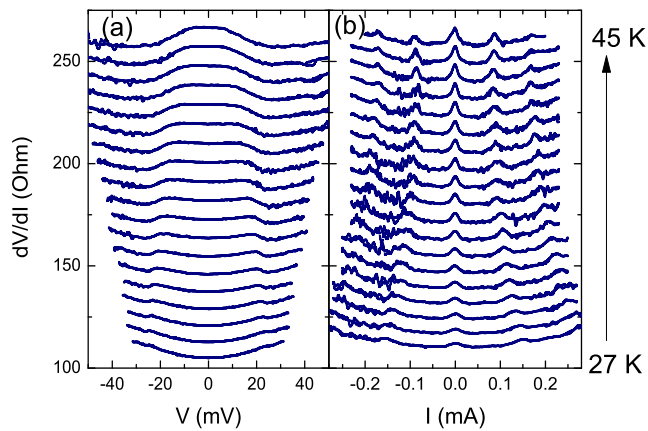}
	\caption{Differential resistance, $dV/dI$, in a NbSe$_3$ 45$^\circ$ cut channel as a function of voltage, $V$, without rf electric field (a) and as a function of current, $I$, under rf electric field with $f=108$ MHz (b) in the temperature range 27-45 K.} \label{F2}
\end{figure}

\section{Results and discussion}

As mentioned above, one of the features of the Hall driven CDW sliding found in Ref. \onlinecite{QM17} is the complete disappearance of  nonlinearity in IVc at temperatures above 35 K. In the present work we measured IVc of a 90$^\circ$-cut channel in a wider temperature range and in different magnetic fields. Fig. \ref{F1} shows IVc measured in magnetic fields 6 T (a), 10 T (b) and 15 T (c) in the temperature range 10-45 K with a step $\Delta T=1$ K. As it can be seen, the threshold effect really disappears with the increase of temperature and is completely absent in a certain range of $T$ (dark yellow curves in Fig. \ref{F1}). However, at further increase of temperature for magnetic fields 6 and 10 T the nonlinearity appears again although with a much smaller magnitude. Such a strange, at first sight, reentrance behavior can be understood taking into account results of Ref. \onlinecite{Hall09} where a field induced change of the Hall voltage sign was observed in NbSe$_3$ in a wide temperature range starting from the lowest ones. With the increase of temperature this zero crossing point moves continuously to higher magnetic field, thus being both magnetic field and temperature dependent. Thus, the effect observed in Ref. \onlinecite{QM17} and attributed to the action of the Hall voltage, must indeed disappear near zero crossing point where no Hall voltage is generated by the normal current. According to Refs. \onlinecite{Hall09,Hall08,Coleman90} these zero crossings correspond to temperatures 24-26, 30-33 and 39-42 K for magnetic fields 6, 10 and 15 T correspondingly. The IV curves corresponding to 25 K at 6 T; 31 K at 10 T and 41 K at 15 T are indicated by the red color in Fig. \ref{F1}. Evidently, in some finite temperature range close to these points, the CDW sliding cannot start because the Hall electric field is below the required threshold. Just this behavior is observed in the present experiments. IVc curves with the dark yellow color in Fig. \ref{F1} indicate the temperature range corresponding to this condition. But at higher temperatures above this zero crossing of the Hall voltage, a finite Hall voltage is recovered. The threshold effect is detected again as seen in Fig. \ref{F1} at 6 and 10 T up to 45 K (blue curves). These results provide the direct confirmation of the Hall effect nature of the CDW sliding in this geometry.

A very unusual feature of the described experimental configuration is the spatially limited sliding region (a few microns): the CDW current should terminate at the edges of the channel and to be converted to the current of normal carriers, most probably by means of phase slippage. To check the possibility of coexistence of the two types of CDW motion we have studied structures with the channel cut at 45$^\circ$ relative to the chain direction. In this case the normal current and hence the driving electric field have components along both directions $b$-axis and $c$-axis.

Fig. \ref{F2} shows differential IVs of such type of structure in zero magnetic field without rf electric field (a) and under rf electric field with $f=108$ MHz (b) in the temperature range 27-45 K. We observe all characteristic features of the conventional collective CDW motion: the resistance decreases sharply at a certain electric field (Fig. \ref{F2} (a)) and pronounced Shapiro steps appear under application of rf electric field with a frequency $f=108$ MHz (Fig. \ref{F2} (b)). The magnitude of nonlinear resistance drop is much smaller compared with the configuration when the external voltage is applied directly along the CDW chains. This fact is not surprising because for the 45$^\circ$-cut channels the main contribution to the voltage drop comes from the normal current across the CDW chains because of the relatively large anisotropy of conductivities $\sigma_b/\sigma_c\approx20$ \cite{OngBrill78}. The CDW transport might be peculiar because the collective current must be terminated at chains edges at the channel boundaries and transformed to normal carriers through phase slippage processes similarly to Hall effect driven sliding. Phase slip processes in submicron distance in NbSe$_3$ have been already reported \cite{Mantel00,VDZant01}.

\begin{figure}[t]
	\includegraphics[width=8.5cm]{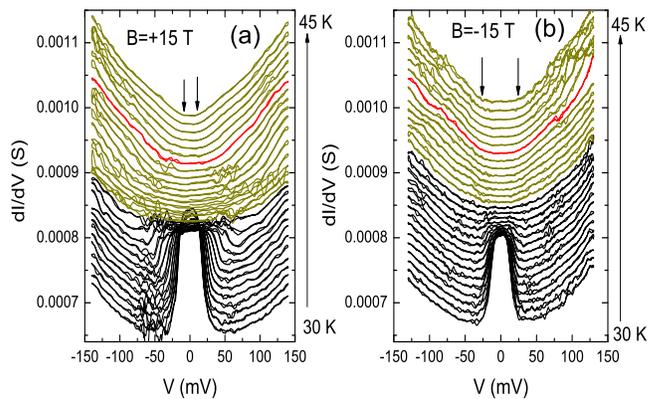}
	\caption{Differential conductance, $dI/dV$ in a NbSe$_3$ 45$^\circ$ cut channel as a function of voltage, $V$, in magnetic field $B=+15$ T (a) and $B=-15$ T at  temperatures from $T=30$ K up to 45 K (from bottom to top) with a step $\Delta T=0.5$ K for the sample shown in Fig.\ref{F2} Arrows indicate the threshold voltage at high temperature, larger for $B=-15$ T for which the components of Hall and external electric fields are opposite in contrast to $B++15$ T when both components add. Note the opposite situation at low temperature consecutive to the change of the sign of the Hall voltage.}
	\label{F3}
\end{figure}
The next step was to study the electronic transport in such type of structure under application of a magnetic field from low temperatures to high temperatures. The idea was to observe both types of sliding: a) Hall driven and b) under application of an external voltage. A curious expectation is that the device properties will essentially depend on the direction of magnetic field ($+B$) or ($-B$). Depending on the sign of $B$, the on-chain components of the Hall and the normal electric field will interfere either positively or negatively, even expecting their amusing canceling.

\begin{figure*}[t]
	\includegraphics[width=0.3\textwidth,height=5.5cm]{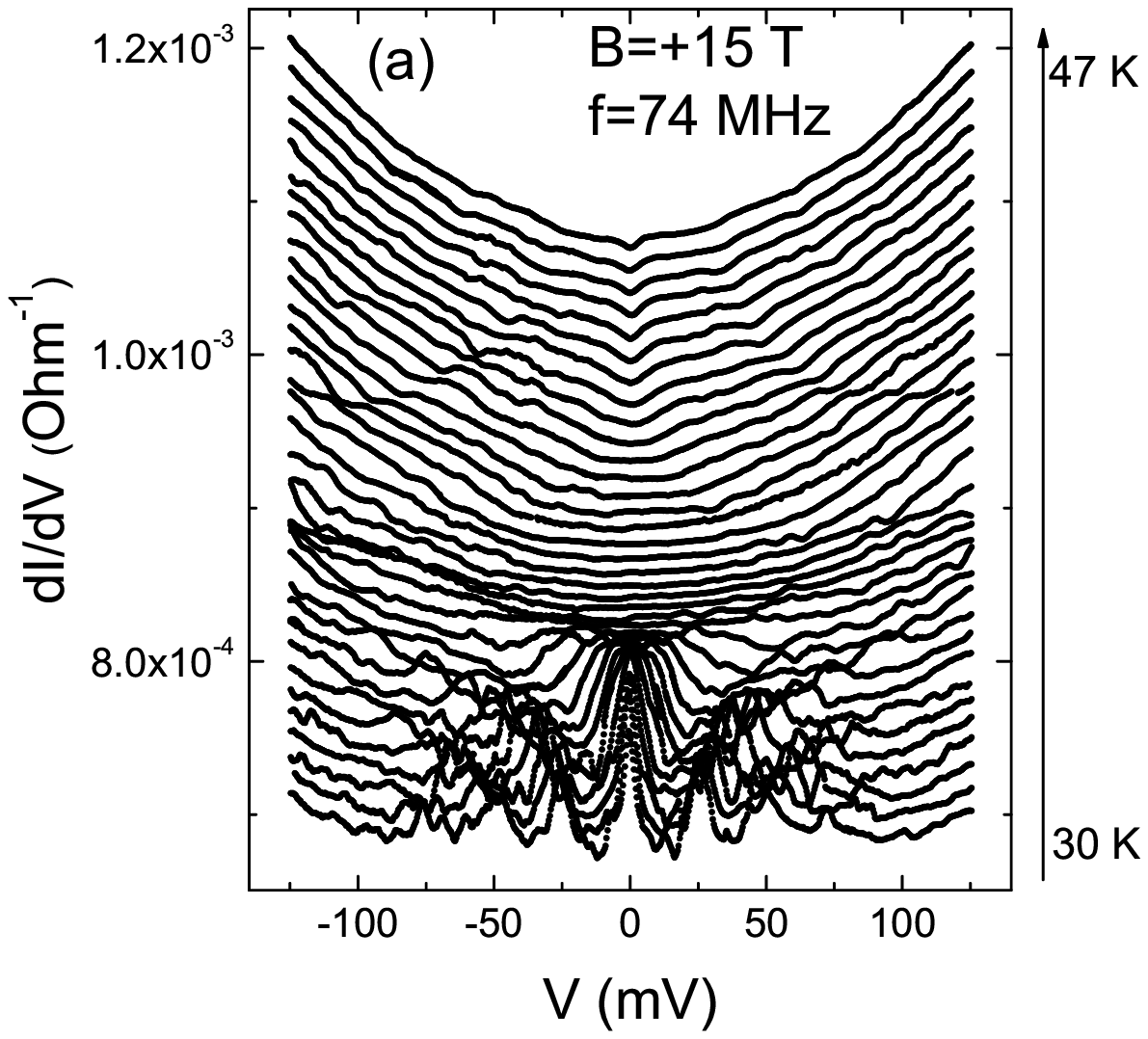} \ \
	\includegraphics[width=0.3\textwidth,height=5.5cm]{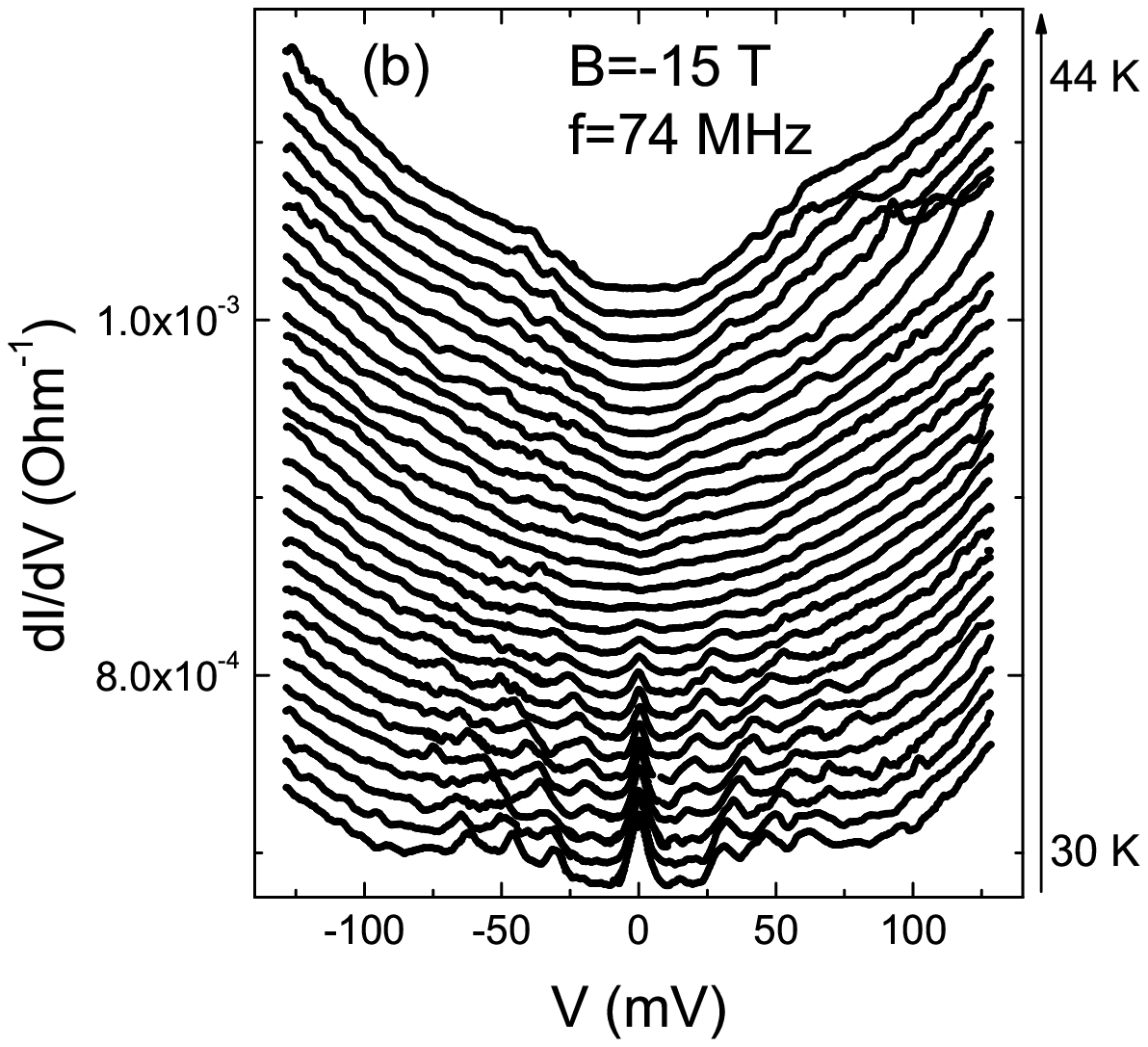} \ \
	\includegraphics[width=0.3\textwidth,height=5.5cm]{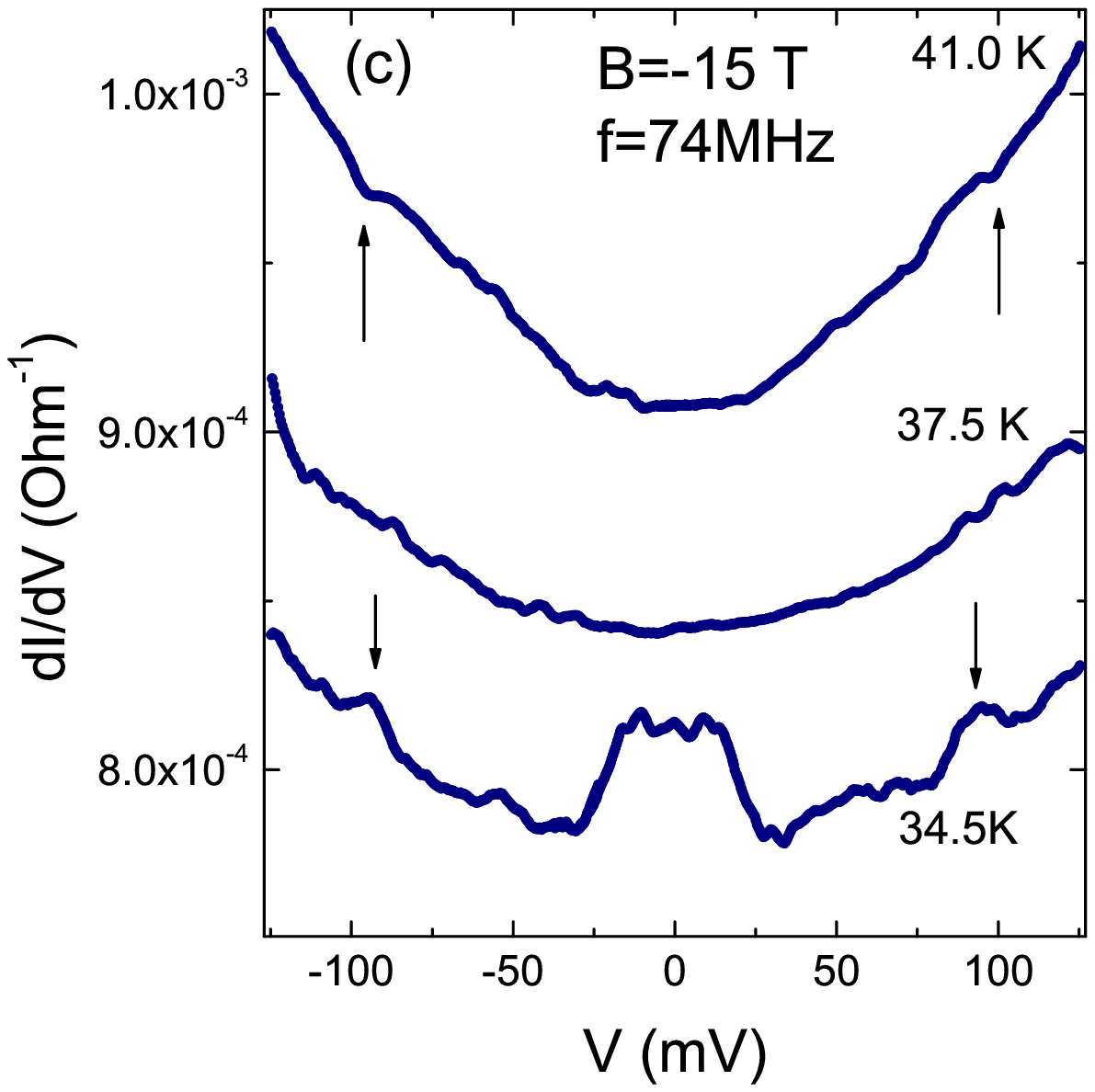}
	\caption{Differential conductance, $dI/dV$ in a NbSe$_3$ 45$^\circ$-cut channel, as a function of voltage, $V$, under rf electric field with $f=74$MHz in magnetic field $B=+15$ T (a); $B=-15$ T (b) at different temperatures for the sample shown in Fig.\ref{F2}; (c) selective $dI/dV$ curves showing the transformation of Hall driven sliding at $T=34.5$ K (peaks of $dI/dV$) to conventional sliding at $T=41$ K (dips of $dI/dV$).} 
	\label{F4}
\end{figure*}

Fig. \ref{F3} demonstrates the temperature evolution of differential IVs of the 45$^\circ$-cut structure in magnetic fields $+15$ T (Fig. \ref{F3} (a)) and $-15$ T (Fig. \ref{F3} (b)). At low temperatures the Hall electric field dominates over the external one due to the very large value of the Hall constant \cite{QM17}. So, for both directions of magnetic field ($+B$) and ($-B$) we observe qualitatively the similar effect of the CDW sliding driven by the Hall electric field as for 90$^\circ$-cut channels: the conductivity drops down at a certain threshold voltage. In the ($+B$) configuration the external electric field is opposite to the Hall field at low $T$ and the threshold electric field for the CDW sliding is larger with respect to the ($-B$) configuration in which both electric fields add (black curves in Fig. \ref{F3}). 

With increase $T$, the Hall electric field decreases because of the decreasing of the Hall constant till a temperature range where the decrease of conductivity due to the Hall voltage driven sliding is transformed in the increase of conductivity due to the conventional current-driven sliding when the component of the external electric field along the chains becomes larger than the corresponding component of the Hall electric field. These curves are indicated by dark yellow colour in Fig. \ref{F3}. The relative sign of the $b$-axis component of the external electric field and the Hall electric field is changed at temperatures above the zero crossing Hall effect (red color in Fig. \ref{F3}). For positive magnetic field the components of Hall and external electric fields add and are opposite for negative magnetic field. As a result, at high temperature the threshold field is smaller for $+B$ and larger for $-B$ as indicated by arrows for $T=$45 K.

It was shown in Ref. \onlinecite{QM17} that under application of an rf electric field in high magnetic field Shapiro steps appear as peaks in $dI/dV(V)$ which are nearly {\it equidistant in voltage} in contrast to conventional CDW sliding in which Shapiro steps appear as a serious of multiple dips {\it equidistant in current}. The transformation of the Hall driven sliding to the conventional one is demonstrated again in Fig. \ref{F4} which shows a set of differential IVc at $B=+15$ T (a) and $B=-15$ T (b) under rf electric field with $f=74$ MHz at different $T$ above 30 K. In both cases at low $T<35$ K we have observed Shapiro steps in the form of peaks in $dI/dV(V)$ as it was observed in the Hall effect driven sliding. Similarly to the data shown in Fig. \ref{F3} (a) for $B=+15$ T, the sliding is interrupted at the temperature of the compensation between opposite Hall electric field and external electric field along the CDW chains. Above this temperature the external electric field along the chains becomes dominant and the Shapiro steps appears again but in the form of minimum of  $dI/dV$ as in the case of conventional sliding. In Fig. \ref{F4} (c) we have drawn a selection of $dI/dV$ curves corresponding to this transformation of the Hall driven ($T=34.5$ K) to the conventional ($T=41$ K) sliding. For $B=-15$ T the temperature corresponding to zero electric field along the CDW chains is much higher than the point of inversion of the sign of the main type of carriers. As a result, the Shapiro steps structure (maxima of conductivity) observed under the Hall effect sliding transforms to conventional Shapiro steps structure (minima of conductivity) continuously when the external electric field along the chains becomes larger than the Hall electric field along the chains.

In conclusion, we have shown that the CDW in NbSe$_3$ can slide being driven by the Hall voltage below the CDW transition temperature, except in the temperature range where the Hall constant crosses zero. These results call for an extension of the theoretical model of Ref. \onlinecite{QM17} in which the remnant normal carriers were expected to be in extreme quantum limit and actually only one kind of carriers was considered. With a narrow channel cut at  45$^\circ$ relative to the chain direction, we have measured the interplay of the Hall driven sliding at low temperature where the Hall constant is large to the conventional sliding at high temperature where the component of the external electric field along the chains becomes larger.

\section{Acknowledgments}

We are thankful to S.A.~Brazovskii for useful discussion. The authors acknowledge the support of the Laboratoirie d'excellence LANEF (ANR-10-LABX-51-01) and of the LNCMI, a member of the European Magnetic Field Laboratory (EMFL). The work has been supported by Russian State Fund for the Basic Research (No. 18-02-00295-a). A.V.F. and A.P.O. thank State assignment IRE RAS.


\begin{thebibliography}{99}
	
\bibitem{Frohlich54}
H. Fr\"{o}hlich, Proc. Roy. Sos. {\bf A223}, 296 (1954).

\bibitem{Fleming79} 
R. M. Fleming and C. C. Grimes, Phys. Rev. Lett. \textbf{42}, 1423 (1979).

\bibitem{Gruner}
G. Gr{\"u}ner, \textit{Density Waves in Solids} (Cambridge, Mass. : Perseus Pub., 2000.), L. Gor'kov  and G. Gr\"uner  \textit
{Charge Density Waves in Solids} (Amsterdam: Elsevier Science, 1989).

\bibitem{Monceau12} 
P. Monceau, Advances in Physics \textbf{61}, 325 (2012).

\bibitem{Bardeen74} 
David Allender, J. W. Bray, and John Bardeen, Phys. Rev. B. \textbf{9}, 119 (1974).

\bibitem{Bardeen85} 
John Bardeen, Phys. Rev. Lett. \textbf{55}, 1010 (1985).

\bibitem{Bardeen89} 
John Bardeen, Phys. Rev. B. \textbf{39}, 3528 (1989).

\bibitem{Thorne88} 
R. E. Thorne, J. S. Hubacek, W. G. Lyons, J. W. Lyding, and J. R. Tucker, Phys. Rev. B. \textbf{37}, 10055 (1988).

\bibitem{Zettl84} 
R. P. Hall and A. Zettl, Phys. Rev. B. \textbf{30}, 2279 (1988).

\bibitem{QM17} 
Andrey P. Orlov, Aleksander A. Sinchenko, Pierre Monceau, Serguei Brazovskii and Yuri I. Latyshev, npj Quantum Materials  61 (2017).

\bibitem{Latyshev11} 
Yu. I. Latyshev, A. P. Orlov, P. Monceau, JETP Lett., \textbf{93}, 101 (2011).

\bibitem{Hall09} 
A.A. Sinchenko, R.V. Chernikov, A.A. Ivanov, P. Monceau, Th. Crozes and S.A. Brazovskii, J. Phys. Condens. Matter. \textbf{21}, 435601 (2009).

\bibitem{Hall08} 
A.A. Sinchenko, Yu.I. Latyshev, A.P. Orlov, A.A. Ivanov, and P. Monceau, Eur. Phys. J. B \textbf{63}, 199 (2008).

\bibitem{Coleman90} 
R.V. Coleman, M.P. Everson, Hao-An Lu, A. Johnson, L.M. Falicov, Phys. Rev. B \textbf{41}, 460 (1990)

\bibitem{OngBrill78} 
N. P. Ong and J.W. Brill, Phys. Rev. B \textbf{18}, 5265 (1978).

\bibitem{Mantel00} 
O. C. Mantel, F. Chalin, C. Dekker, H. S. J. van der Zant, Yu. I. Latyshev, B. Pannetier, and P. Monceau, Phys. Rev. Lett. \textbf{84}, 538 (2000).

\bibitem{VDZant01} 
H. S. J. van der Zant, E. Slot, S. V. Zaitsev-Zotov, and S. N. Artemenko, Phys. Rev. Lett. \textbf{87}, 126401 (2001).


\end{thebibliography}
\end{document}